\documentclass[aps,prl,twocolumn,superscriptaddress,showpacs,floatfix,preprintnumbers,nofootinbib]{revtex4-2}

\usepackage[utf8]{inputenc}

\usepackage{cancel}

\usepackage[dvipsnames]{xcolor}      
\definecolor{lcolor}{rgb}{0.5,0,0}
\definecolor{citcolor}{rgb}{0,0.0,1}

\usepackage[breaklinks,colorlinks,urlcolor=blue,citecolor=citcolor,linkcolor=lcolor,linktoc=all]{hyperref}
\usepackage{color}
\usepackage{graphicx}	
\graphicspath{{./figures/}}
\usepackage[utf8]{inputenc}
\usepackage{amsmath} 
\usepackage{amssymb}
\usepackage[ragged]{footmisc}
\usepackage{slashed} 
\usepackage{float}
\usepackage[capitalise]{cleveref}

\makeatletter
\g@addto@macro\bfseries{\boldmath}
\makeatother

\usepackage{tikz}
\usepackage[customcolors]{hf-tikz}
\usepackage{mciteplus}

\usetikzlibrary{arrows,cd,shapes,decorations.pathmorphing,decorations.markings,shadings}
\tikzset{
  big arrow/.style={
    decoration={markings,mark=at position 1 with {\arrow[scale=4,#1]{>}}},
    postaction={decorate},
    shorten >=0.4pt},
  big arrow/.default=blue}

\newcommand{\bbone}{\text{\usefont{U}{bbold}{m}{n}1}}
\MakeRobust{\bbone}

\newcommand{\Lh}{\Lambda_{\text{h}}}
\newcommand{\Lbar}{\overline{\Lambda}}
\newcommand{\mE}{m_{\text{E}}}

\newcommand{\muB}{\mu_\text{B}}
\newcommand{\muQ}{\mu_q}

\newcommand{\nc}{N_c}
\newcommand{\ca}{C_A}
\newcommand{\cf}{C_F}
\newcommand{\da}{d_A}
\newcommand{\nf}{N_f}

\renewcommand{\epsilon}{\varepsilon}

\newcommand{\ud}{\mathrm{d}}

\newcommand{\as}{\alpha_s}

\newcommand{\gamE}{\gamma_{\text{E}}}
\newcommand{\tr}[1]{\operatorname{Tr}\left \{ #1 \right \} }

\newcommand{\xih}{\hat\xi}
\newcommand{\Ae}{\mathsf{A}}
\newcommand{\Be}{\mathsf{B}}
\newcommand{\Le}{\mathsf{L}}
\newcommand{\He}{\mathsf{H}}

\newcommand{\Je}{\mathsf{J}}

\newcommand{\normf}{\lambda}

\newcommand{\lo}{\text{LO}}
\usepackage[acronym]{glossaries}
\newacronym{LO}{LO}{leading-order}

\newacronym{QM}{QM}{quark matter}

\newacronym{QCD}{QCD}{quantum chromodynamics}

\newacronym{HTL}{HTL}{Hard-Thermal-Loop}

\newacronym{UV}{UV}{ultraviolet}

\newacronym{IR}{IR}{infrared}

\newcommand{\MSbar}{$\overline{\text{MS}}$}

\begin{document}

\title{\texorpdfstring{Equation of State of Cold Quark Matter to $O(\alpha_s^3 \ln \alpha_s)$}{Equation of State of Cold Quark Matter to O(as\^{}3 ln as)}}

\preprint{HIP-2023-10/TH, TUM-EFT 181/23}
\author{Tyler Gorda}
\email{tyler.gorda@physik.uni-frankfurt.de}
\affiliation{Technische Universit\"{a}t Darmstadt, Department of Physics, 64289 Darmstadt, Germany}
\affiliation{ExtreMe Matter Institute EMMI, GSI Helmholtzzentrum f\"ur Schwerionenforschung GmbH, 64291 Darmstadt, Germany}
\author{Risto Paatelainen}
\email{risto.paatelainen@helsinki.fi}
\affiliation{Department of Physics and Helsinki Institute of Physics, P.O.~Box 64, FI-00014 University of Helsinki, Finland}
\author{Saga S\"appi}
\email{saga.saeppi@tum.de}
\affiliation{TUM Physik-Department, Technische Universität München, James-Franck-Str.~1, 85748 Garching, Germany}
\affiliation{Excellence Cluster ORIGINS, Boltzmannstrasse 2, 85748 Garching, Germany}
\author{Kaapo Seppänen}
\email{kaapo.seppanen@helsinki.fi}
\affiliation{Department of Physics and Helsinki Institute of Physics, P.O.~Box 64, FI-00014 University of Helsinki, Finland}

\begin{abstract}
Accurately understanding the equation of state (EOS) of high-density, zero-temperature quark matter plays an essential role in constraining the behavior of dense strongly interacting matter inside the cores of neutron stars. In this Letter, we study the weak-coupling expansion of the EOS of cold quark matter and derive the complete, gauge-invariant contributions from the long-wavelength, dynamically screened gluonic sector at next-to-next-to-next-to-leading order (N3LO) in the strong coupling constant $\alpha_s$. This elevates the EOS result to the $O(\alpha_s^3 \ln \alpha_s)$ level, leaving only one unknown constant from the unscreened sector at N3LO, and places it on par with its high-temperature counterpart from 2003.
\end{abstract}

\maketitle

\emph{Introduction.}---%
A proper understanding of the thermodynamics of dense strongly interacting matter is an outstanding problem in theoretical physics. This is in large part due to the infamous sign problem of lattice field theory \cite{Hasenfratz:1983ba,Kogut:1983ia,Karsch:1985cb,deForcrand:2009zkb,Nagata:2021ugx}, which renders nonperturbative lattice Monte Carlo techniques largely inapplicable in the region of large baryon chemical potential $\muB$ and low temperatures $T$ in quantum chromodynamics (QCD). 

At densities above $40$ times the nuclear saturation density, $n_0 \approx 0.16$~fm$^{-3}$, a
perturbative weak-coupling expansion in the strong coupling constant $\alpha_s$ becomes applicable within the fundamental theory of QCD, due to asymptotic freedom. In this regime, it becomes possible to study the thermodynamics of cold (zero-temperature) quark matter (QM) directly using well-established thermal-field-theory tools \cite{Kapusta:2006pm,Bellac:2011kqa,Laine:2016hma,Ghiglieri:2020dpq}. The equation of state (EOS) of high-density cold QM has in recent years received increasing attention as a robust high-density constraint \cite{Komoltsev:2021jzg} to be used when performing neutron-star EOS inference at lower densities \cite{Kurkela:2014vha,Annala:2017llu,Most:2018hfd,Altiparmak:2022bke,Annala:2019puf,Annala:2021gom,Gorda:2022jvk,Somasundaram:2022ztm,Jiang:2022tps,Fujimoto:2022ohj,Marczenko:2022jhl,Gorda:2022lsk,Sorensen:2023zkk,Takatsy:2023xzf, Essick:2023fso,Brandes:2023hma,Mroczek:2023zxo}. In addition to this phenomenological application, there is great theoretical interest to study high-density cold QM, due to the rich physics arising from the dynamical screening of long-wavelength chromoelectric and chromomagnetic fields. These screening effects necessitate the development of an effective field theory of the long-wavelength gluonic modes, which goes beyond a fixed loop order in the weak-coupling expansion.

Such screening also occurs in a high-temperature quark-gluon plasma. However, at high temperatures, low-energy chromoelectric and chromomagnetic gluons can additionally receive large thermal occupation numbers $\sim 1/\alpha_s^{1/2}$ or $\sim 1/\alpha_s$, respectively \cite{Blaizot:2001nr}. These large occupation numbers lead to poor perturbative convergence within the long-wavelength chromoelectric sector of QCD and necessitate a nonperturbative treatment of the chromomagnetic sector at high temperatures. Crucially, however, such a long-wavelength ``Bose enhancement'' is absent within high-density unpaired cold QM, and hence the thermodynamics of such a system \emph{always} remains a perturbative problem. Furthermore, we also emphasize that this conclusion is not modified by any possible color superconducting phases that may exist at high densities, since the corrections to bulk thermodynamics from these effects are exponentially suppressed $\sim \exp(-\#/\alpha_s^{1/2})$ \cite{Son:1998uk,Malekzadeh:2006ud,Alford:2007xm}.

Since long-wavelength gluons are not Bose enhanced in cold QM, one could, in principle, extend the results for the EOS to even lower densities by tackling increasingly higher-order perturbative computations---something which could lead to dramatic improvements within the aforementioned neutron-star EOS-inference setups. In this Letter, we compute the full EOS contributions from the screened gluonic sector at next-to-next-to-next-to-leading order (N3LO) in $\alpha_s$ in unpaired cold QM, bringing the result to the $O(\alpha_s^3 \ln \alpha_s)$ level and on par with its high-temperature counterpart from 2003 \cite{Kajantie:2002wa}. This is achieved by deriving the next-to-leading order (NLO) screening corrections to long-wavelength gluon propagation in cold QM, generalizing the recent results in \cite{Gorda:2023zwy,Ekstedt:2023oqb} to zero temperature. We find that our result shows very small renormalization-scale dependence for $\muB$ where it is convergent, suggesting that the screened gluonic sector is under good perturbative control at high densities.

\emph{Structure of the weak-coupling expansion of the EOS.}---%
In the context of cold QM with vanishing quark masses in the grand canonical ensemble, the EOS is given by the free energy density $\Omega = -p$, where $p$ is pressure, as a function of the quark and lepton chemical potentials. In astrophysical environments, the conditions of charge-neutral, three-flavor quark matter in equilibrium under the weak interactions (beta equilibrium) are the most relevant. These conditions reduce the EOS to a function of a single quantity $\muB$. Up to and including NLO, the EOS, which we henceforth identify as $p(\muB)$, is sensitive only to ``hard'' quark and gluonic corrections from momenta $K \sim \muB$, since the low-momentum quarks are Pauli blocked and the screened gluonic sector is phase-space suppressed: $\int_{K < \alpha_s^{1/2} \muB} d^4 K \simeq \alpha_s^2 \muB^4.$ However, beginning at next-to-next-to-leading order (N2LO) in $\alpha_s$, which was computed in 1977 in the context of cold QM \cite{Freedman:1976dm,Freedman:1976ub}, the pressure becomes sensitive to long-wavelength, screened gluonic fluctuations. These field modes can be described via the hard thermal loop (HTL) effective theory \cite{Braaten:1989mz,Blaizot:2001nr} (or hard dense loop (HDL) \cite{Manuel:1995td,Ipp:2003cj,Ipp:2006ij}), which captures the physics of the ``soft'' momentum scale $\mE \sim \as^{1/2} \muB$, corresponding to chromoelectric screening. At N2LO, because of the phase-space suppression, self-interactions between these screened gluons do not yet contribute. At the next order in $\alpha_s$, however, interactions involving these screened fluctuations must be included. In particular, there arise contributions from self-interactions among the fluctuating soft modes and also interactions between fluctuating soft and hard modes. The pressure correction arising from the self-interactions at N3LO has been computed within the leading-order HTL theory  \cite{Gorda:2021znl,Gorda:2021kme}, while the pressure correction arising from interactions between the soft and hard modes involves corrections to the leading-order HTL theory. These corrections to high-density HTL have been calculated within QED \cite{Carignano:2017ovz,Carignano:2019ofj,Gorda:2022fci} at zero and high temperatures, and recently within QCD \cite{Ekstedt:2023oqb,Gorda:2023zwy} at high temperatures.

In total therefore, the pressure of cold QM at N3LO can be written in the form 
\begin{equation}\label{eq:pressectors}
p_\mathrm{N3LO} = p^s + p^m + p^h,
\end{equation}
where the three terms on the right-hand side come from the different combinations of soft and hard momentum scales. The first term $p^s$ is the purely soft contribution arising from self-interactions between screened gluon field modes. The second term $p^m$, dubbed the mixed contribution, arises from the corrections within the HTL theory. 
Finally, in addition to soft and mixed terms, there are also fully hard contributions $p^h$, entering the expansion via four-loop vacuum graphs in full QCD (see \cite{Gorda:2021kme} for a list). In this work, we compute the screened correction $p^m$ from the mixed sector.

The screened gluonic modes lead to nonanalyticities in the pressure, arising from logarithms of the ratio of the soft and hard scales $\ln(\mE / \muB)$. To N3LO, the pressure $p$ of cold and dense, beta-equilibrated three-color, three-flavor ($\nc=\nf=3$) unpaired QM with massless quarks can be cast into the following form \cite{Baym:1975va,Chapline:1976gq,Freedman:1976ub,Gorda:2018gpy}, (see \cite{supp_mat} for the general $\nc$ and $\nf$ expressions)
\begin{widetext}
\begin{equation}
\label{eq:QMpressure}
\frac{p}{p_{\text{free}}}  \simeq 1 - 2\left (\frac{\as}{\pi}\right ) - 3\left (\frac{\as}{\pi}\right )^2 \biggl [\ln \left (3\frac{\as}{\pi}\right )+3\ln X +5.0021 \biggr ] 
+ 9\left (\frac{\as}{\pi}\right )^3\biggl [c_{3,2} \ln^2 \left (3\frac{\as}{\pi}\right ) + c_{3,1}(X)\ln \left (3\frac{\as}{\pi}\right ) + c_{3,0}(X)\biggr ]  
\end{equation}
\end{widetext} 
where $p_{\text{free}} = 3(\muB/3)^4/4\pi^2$ is the pressure of a free Fermi gas of quarks in beta equilibrium, $\as = \as(\Lbar)$ is the renormalized strong coupling constant in the \MSbar\ scheme at the renormalization scale $\Lbar$, and $X \equiv 3\Lbar/(2\muB)$ is a quantity which should be taken to be $O(1)$ to minimize large logarithms. We have written logarithms of $\as$ by grouping them in the natural expansion parameter $N_f \as / \pi$, and set $N_f=3$. The coefficient $c_{3,2}$ of the leading logarithm was originally determined in \cite{Gorda:2018gpy}. Recently, an all-order leading-logarithm resummation was conducted in \cite{Fernandez:2021jfr}, using this term as input.

In this work, we conclusively determine the coefficient $c_{3,1}(X)$ of the next-to-leading logarithm. It should be emphasized that this term is a well-defined and independent coefficient in the perturbative series. While the coefficient is now fully determined, the parts of the $c_{3,1}(X)$-contribution proportional to $\ln X$ could already be inferred from renormalization group invariance using lower-order results. Likewise, given this newly determined contribution, all parts of $c_{3,0}(X)$ proportional to $\ln^k X$ can now be inferred with similar arguments, thus resulting in the entire $p_{\text{N3LO}}$ except for one constant from the hard sector.

\emph{Overview of the calculation.}---%
The mixed contributions correspond to the following diagrams stemming from the  classification of \cite{Gorda:2021kme}
\begin{widetext}
\begin{align}
\label{eq:mixeddiags}
    p^{m}  =& \left(
    \raisebox{-0.44\height}{\includegraphics[height=1.0cm,page=1]{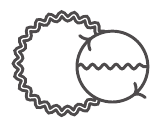}} + \raisebox{-0.44\height}{\includegraphics[height=1.0cm,page=2]{figures/mixed-diags.pdf}} + \raisebox{-0.44\height}{\includegraphics[height=1.0cm,page=3]{figures/mixed-diags.pdf}} + \raisebox{-0.44\height}{\includegraphics[height=1.0cm,page=4]{figures/mixed-diags.pdf}} +\raisebox{-0.44\height}{\includegraphics[height=1.0cm,page=5]{figures/mixed-diags.pdf}} \right)  + \left(\raisebox{-0.44\height}{\includegraphics[height=1.0cm,page=6]{figures/mixed-diags.pdf}} + \raisebox{-0.44\height}{\includegraphics[height=1.0cm,page=7]{figures/mixed-diags.pdf}} + \raisebox{-0.44\height}{\includegraphics[height=1.0cm,page=8]{figures/mixed-diags.pdf}} + \raisebox{-0.44\height}{\includegraphics[height=1.0cm,page=9]{figures/mixed-diags.pdf}} \right) \nonumber \\
=& \raisebox{-0.44\height}{\includegraphics[height=1.2cm,page=10]{figures/mixed-diags.pdf}} + \raisebox{-0.44\height}{\includegraphics[height=1.2cm,page=11]{figures/mixed-diags.pdf}} = - \frac{\as\mE^2\da}{8\pi} \int_K \tr {G_\lo(K) \left[ \Pi^{2,\text{HTL}}(K)+ \frac{K^2}{\mE^2} \Pi^{1,\text{Pow}}(K)\right]}.
\end{align}
\end{widetext} 
The double wavy lines correspond to soft, HTL-resummed gluons; the wavy, solid, and dotted lines correspond to hard, unresummed gluons, quarks, and ghosts, respectively; and the trace is over the suppressed Lorentz indices. Furthermore, a sum over the direction of fermionic flow is implied. In \cref{eq:mixeddiags}, the two shaded blobs and the respective $\Pi$'s correspond to the two distinct NLO soft gluon self-energies at zero temperature: namely, the two-loop corrections given in the first pair of parentheses and the power corrections, given by the $O(K^2)$ term in a small-$K$ expansion, given in the second. The $G_\lo$ is the standard one-loop resummed HTL propagator \cite{Braaten:1989mz,Manuel:1995td}, $\da=N_c^2-1$, $K$ is the loop momentum associated with the soft gluon, and the $\Pi$'s have been rescaled to be dimensionless and independent of $\as$,  (see details and explicit expressions in \cite{supp_mat}). In dimensional regularization and in the \MSbar\ scheme, the integration measure is defined by $\int_K \equiv \left (e^{\gamE}\Lh^2/(4\pi) \right )^{(4-D)/2}\int \ud^D K/(2\pi)^D$, where $D \equiv 4-2\varepsilon$ and $\Lh$ is the factorization scale associated with the split between the hard and soft modes \cite{Gorda:2021kme, Gorda:2022zyc}. The complete Feynman rules can be found in \cite{Gorda:2021kme}.

Note that $\Pi^{2,\mathrm{HTL}}$ and $\Pi^{1,\mathrm{Pow}}$ entering in \cref{eq:mixeddiags} have been recently computed at high temperature and large $\muB$ in general covariant gauge \cite{Gorda:2023zwy}, where they have been shown to contain $\ln T$ terms that diverge in the zero-temperature limit. These terms arise because the temperature regulates certain infrared (IR) divergences associated with the hard internal gluon lines in the self-energy diagrams. The corresponding zero-temperature self-energies are computed in \cite{supp_mat} in terms of $d$-dimensional integral expressions. The zero-temperature limit is achieved by using the exact integral expressions of these self-energies and isolating divergent bosonic integrals that vanish at exactly zero temperature in dimensional regularization. This procedure effectively converts, after renormalization, the $\ln T$ terms into IR $1/\epsilon$ terms in the strict zero-temperature expressions. We find that the final expressions for the self-energies depend linearly on the gauge parameter $\xi$ in a general covariant gauge, in contrast to the quadratic dependence found at high temperatures in \cite{Gorda:2023zwy}.

With the renormalized self-energies in hand, the mixed contribution to the pressure can then be obtained by computing the integral in \cref{eq:mixeddiags}. The integral over the soft momentum $K$ contains ultraviolet (UV) divergences that arise because the HTL theory differs from full QCD in the UV. These UV divergences must cancel with corresponding IR divergences in the hard theory, contained in $p^h$. This cancellation has been explicitly shown in QED, where the soft sector trivially vanishes, \cite{Gorda:2022zyc}, but due to the added complexity of QCD, we do not show this explicitly---however, the cancellation must occur as long as HTL does indeed correctly describe the soft physics at this order. 

Importantly, upon computing the radial $K$ integral in \cref{eq:mixeddiags}, we find that only the specific combination of self-energies $\Pi^{2,\mathrm{HTL}} - \Pi^{1,\mathrm{HTL}} \Pi^{1,\mathrm{Pow}}$ appears. Here $\Pi^{1,\mathrm{HTL}}$ denotes the standard one-loop HTL self-energy \cite{Braaten:1989mz,Manuel:1995td}, appearing in $G_\mathrm{LO}$. This particular combination is explicitly gauge-invariant, guaranteeing the gauge invariance of the mixed contribution to the pressure. We remark here that this should be the case since the 2-loop HTL pressure corresponding to the soft sector is known to be gauge independent \cite{Andersen:2002ey,Gorda:2021kme}, and the 4-loop hard-sector diagrams are likewise known to be algebraically gauge invariant \cite{Schroder:2002re}. We note also that the above self-energy combination also naturally appears when one rescales the HTL effective Lagrangian including the 2-loop and power corrections to bring the kinetic term to a canonical form \cite{Kajantie:1995dw,Ekstedt:2023oqb}.

\emph{Results and discussion.}---%
Upon computing the renormalized $p^m$ in dimensional regularization (details are given in \cite{supp_mat}, in general $\nc$ and $\nf$), we find it to have a similar form to the $p^s$ computed in \cite{Gorda:2021kme,Gorda:2021znl}, namely,%
\begin{equation}
\begin{split}
\label{eq:p3mixedv1}
p^m  = 
\frac{\as\mE^4\da}{(4\pi)^3} \left(\frac{\mE}{\Lh}\right)^{-2\varepsilon}\left(\frac{\muB/3}{\Lh}\right)^{-2\varepsilon} \\
\times \left(\frac{p^m_{-2}}{(2\varepsilon)^2} + \frac{p^m_{-1}(X)}{2\varepsilon} + p^m_{0}(X) \right),
\end{split}
\end{equation}
where the coefficients $p_i^m$ denote terms in the $\epsilon$ expansion and do not depend on the coupling. The factor $(\mE/\Lh)^{-2\epsilon}$ arises from the integral over the soft-theory loop momentum $K$ in \cref{eq:mixeddiags}, while the factor $(\muB/(3\Lh))^{-2\epsilon}$ arises from the hard-theory calculation leading to the self-energies. 
We find the new coefficients $p^m_i$ in \cref{eq:p3mixedv1} to be
\begin{equation}
\label{eq:p3results}
\begin{split}
p^m_{-2} & = -11; \quad p^m_{-1}(X) = 9\ln X -4.8095, \\
p^m_0(X) & = -\frac{9}{2}\ln^2 X + 2.0598\ln X -5.6316.
\end{split}
\end{equation}
We note that this now confirms through an explicit calculation a prediction made in \cite{Gorda:2021kme}: $p^m_{-2}=-2p^s_{-2}$.

We then combine this result with similar expressions for $p^s$ and $p^h$ in \cref{eq:pressectors} (whose expressions are given in \cite{supp_mat}) and obtain the renormalized result for the full N3LO pressure 
\begin{equation}
\label{eq:p3unknowns}
\begin{split}
&p_\mathrm{N3LO}  = 
\frac{\as\mE^4\da}{(4\pi)^3}\Biggl[p^s_{-2}\ln^2\left(\frac{\muB/3}{\mE}\right) \\
&\,+\left(2p^s_{-1}+p^m_{-1}(X)\right)\ln\left(\frac{\muB/3}{\mE}\right)+p^s_0+p^m_0(X)+p^h_0(X)\Biggr]. 
\end{split}
\end{equation}
As mentioned above, we here assume the intermediate IR and UV divergences between the different sectors to fully cancel, which in turn ensures that the $\Lh$ dependence from the different sectors cancels. 
From renormalization-scale independence of the partial result (see, e.g., \cite{Gorda:2022zyc}), we are additionally able to determine the full $X$ dependence of $p_0^h(X)$, leading to the form 
\begin{equation}
\label{eq:ph0}
\begin{split}    
p^h_0(X) & = -\frac{9}{4} \ln^2 X -26.367 \ln X + c_0.
\end{split}
\end{equation}
Equations~\eqref{eq:p3mixedv1}--\eqref{eq:ph0} are our main result, and they fix the coefficients in the pressure in \cref{eq:QMpressure} to be those given in \cref{tab:pressurecoeffsmain}, with $c_0$ the remaining unknown constant from the hard sector.

\begin{table}[t!]
\setlength\tabcolsep{0pt}
\begin{tabular*}{\linewidth}{@{\quad}l@{\hspace{50pt}}l}
\toprule
    $c_{3,2}$ & $11/12$ \\
    $c_{3,1}$ & $-6.5968(12)-3\ln X$ \\
    $c_{3,0}$ &  $5.1342(48)+\frac{2}{3}c_0-18.284 \ln X-\frac{9}{2}\ln^2 X$ \\
\botrule
\end{tabular*}
    \caption{List of numerical values for the coefficients appearing in \cref{eq:QMpressure}.}
    \label{tab:pressurecoeffsmain}
\end{table}

\begin{figure}[t!]
    \centering \includegraphics[width=0.48\textwidth]{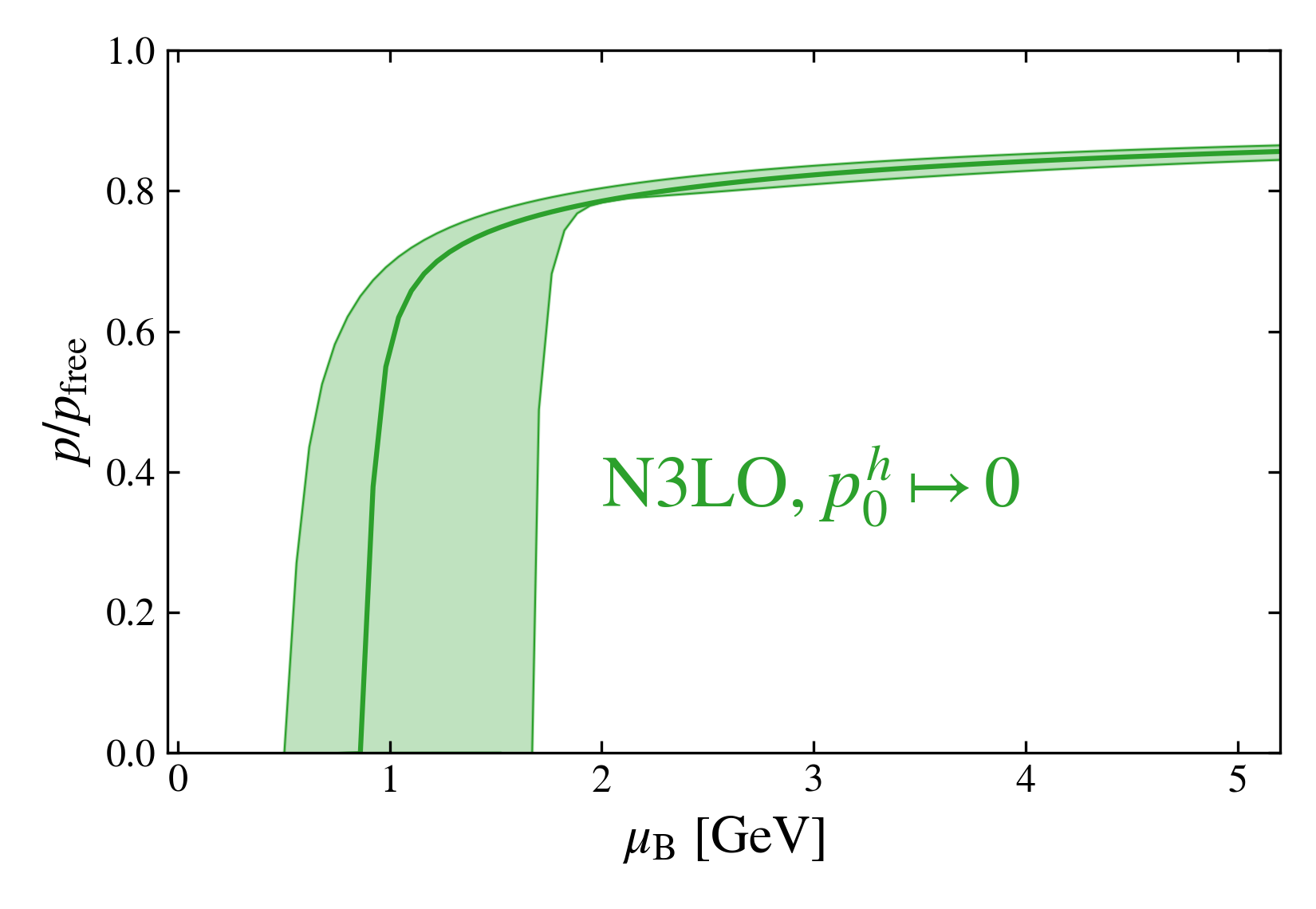}
    \caption{The N3LO pressure normalized by the free pressure as a function of baryon chemical potential, including all contributions but the hard contributions $p^h_0(X)$ in \cref{eq:ph0}. The shaded region shows the usual scale variation of $X \in [1/2,2]$, while the solid line is the central scale choice $X = 1$.}
    \label{fig:p_N3LO_sm}
\end{figure}

In \cref{fig:p_N3LO_sm} we show the partial N3LO pressure, neglecting only the finite hard contribution $p^h_0(X)$ at this order. In the figure, the uncertainty of the truncation of the perturbative series is estimated by varying of $X \in [1/2, 2]$, shown as a shaded uncertainty band. In this and all subsequent figures, we use the three-loop beta function when computing the running $\alpha_s(\Lbar)$. We see from \cref{fig:p_N3LO_sm} that the pressure contribution when including the screened gluonic sector at N3LO is remarkably well converged. In particular, it has nearly vanishing renormalization-scale dependence for all $\muB > 2$~GeV. In fact, we have verified that one obtains a similarly well-converged result when neglecting the hard contribution at N2LO as well. This is consistent with the observation made in \cite{Gorda:2021kme,Gorda:2021znl}: In cold QM, it is the hard terms $p^h(X)$ that drive the inevitable breakdown of perturbation theory, in stark contrast to the high-temperature case where the soft modes are responsible for the breakdown. 

We turn now to an analysis of the full N3LO pressure, which we have now determined up to a single unknown constant from the hard sector $c_0$. Upon substituting in different values, we find that the N3LO pressure strongly depends on $c_0$. An approximate value for this constant can only be determined by computing many hard four-loop diagrams---a project which will take several years of effort. Hence, we are motivated to find an alternative way to estimate its value in order to quantify how well converged we may expect the full N3LO result to be.

Since the perturbative series for the pressure of cold QM is well-behaved up to N2LO, we turn to a Bayesian model to identify which values of $c_0$ are most consistent with lower-order results \cite{Cacciari:2011ze}. In particular, we use the $abc$ model of convergent series as presented in \cite{Duhr:2021mfd} and implemented in the publicly available \texttt{MiHO} code \cite{mihogit}. This Bayesian model, as well as the geometric model from \cite{Bonvini:2020xeo} was recently applied to the high-density perturbative-QCD EOS in \cite{Gorda:2023usm}. These models assume that the perturbative series can be approximated by independent draws from a statistical model of convergent series. Upon conditioning these models using the NLO and N2LO results using Bayes's theorem, they provide posterior distributions for the N3LO pressure as a function of $\muB$ and $X$, quantifying the degree to which a given $p_\mathrm{N3LO}$ is consistent with the lower-order results. We choose to perform the following analysis at a fixed $\muB = 2.6$~GeV, which is the canonical value down to which the N2LO results are typically used \cite{Kurkela:2014vha}. We have checked that our conclusions remain similar if we choose a slightly different matching $\muB$ (see also \cite{Gorda:2023usm}). 

As the N3LO pressure is now a function of a single parameter $c_0$, posterior distributions of $p_\mathrm{N3LO}$ at fixed $X$ can be converted to distributions of $c_0$. To combine the resulting $c_0$ distributions for different $X$ values, we choose to marginalize over the fictitious renormalization parameter $X \in [1/2, 2]$ following the procedure introduced in \cite{Bonvini:2020xeo}. This marginalization weights different values of the renormalization parameter in proportion to how well the $abc$ model is converged at that $X$. This results in larger values of $X$ receiving slightly larger weights, and furthermore leads to a slightly more conservative distribution than would result from the alternative scale-averaging procedure \cite{Duhr:2021mfd} in this case.

\begin{figure*}
    \centering \includegraphics[width=0.48\textwidth]{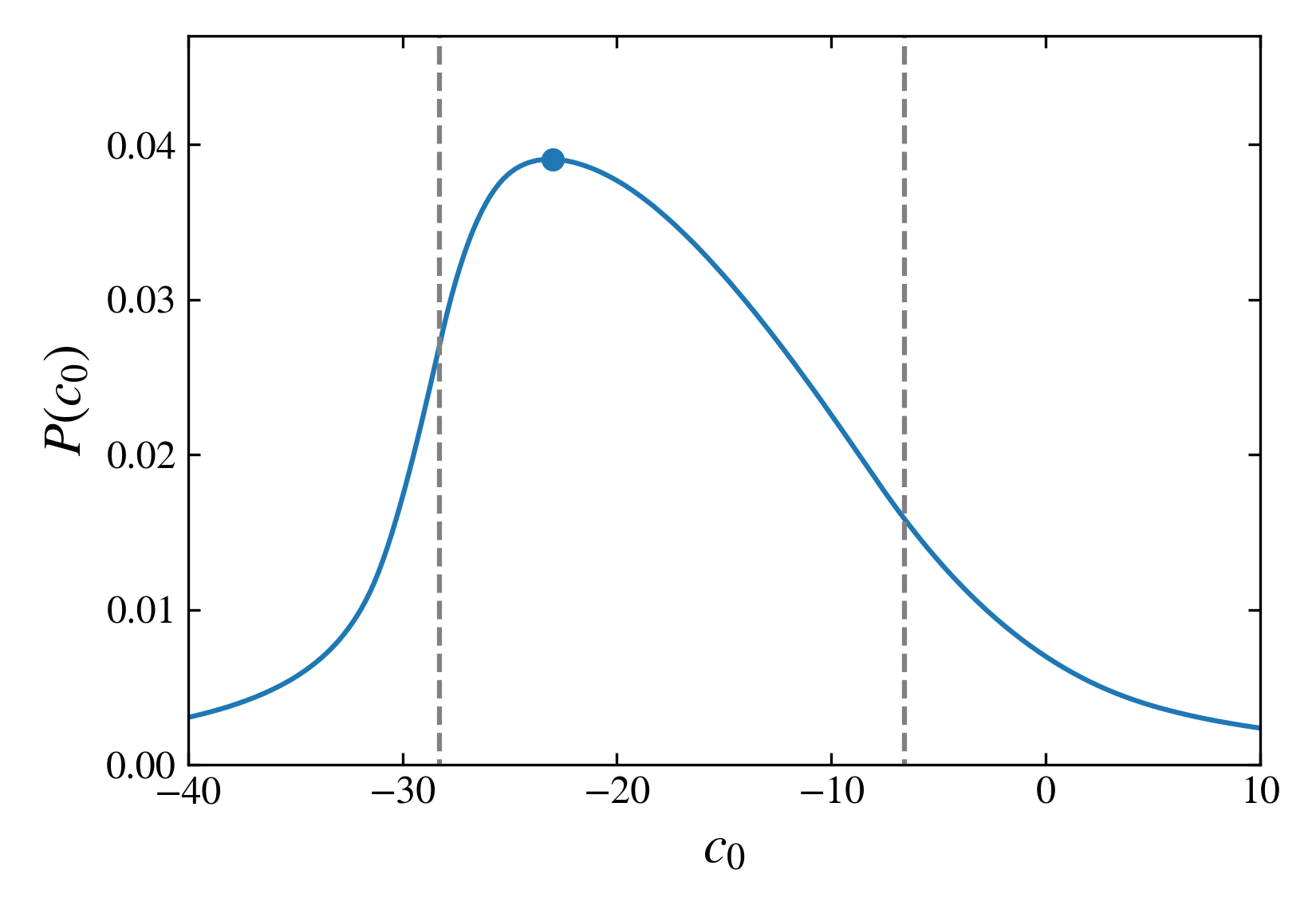}  \includegraphics[width=0.48\textwidth]{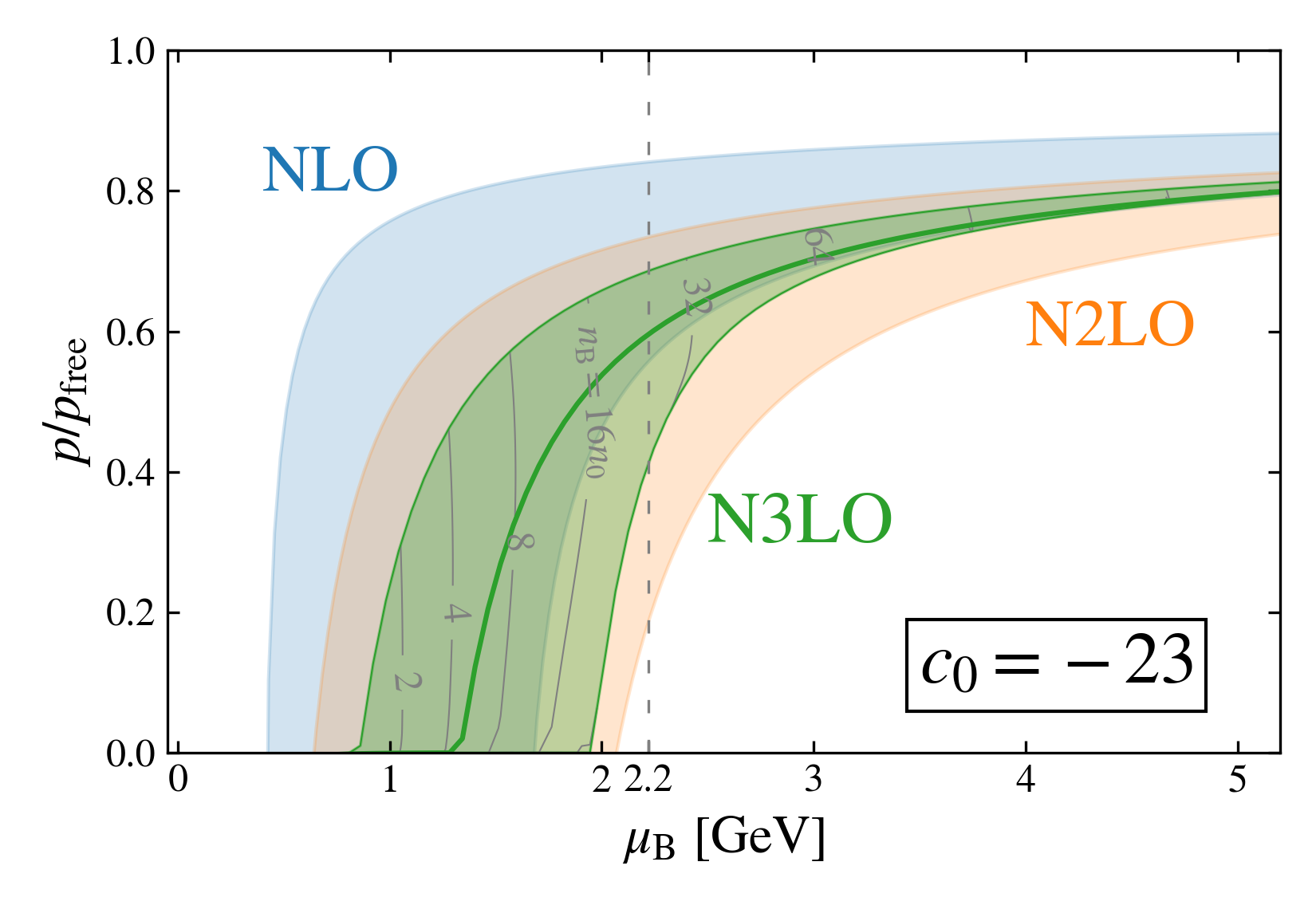}
    \caption{(Left) The posterior distribution of $c_0$ within the $abc$ model after marginalizing over $X \in [1/2, 2]$ \cite{Duhr:2021mfd} at $\muB = 2.6$~GeV (see main text). The dashed lines correspond to the 68\% credible interval, and the solid point is the maximum of the distribution. (Right) The resulting N3LO pressure for most-consistent value $c_0 = -23$. The dashed line at $\muB = 2.2$~GeV denotes the point where the N3LO result possesses $\pm 25\%$ errors, and the gray lines denote contours of constant number density.}
    \label{fig:p_N3LO_predict}
\end{figure*}

The above analysis leads to the probability distribution $P(c_0)$ shown in the left panel of \cref{fig:p_N3LO_predict}. The distribution is rather broad, with a 68\% credible interval corresponding to $c_0 \in [-29, -6]$, denoted in \cref{fig:p_N3LO_predict} by the dashed gray lines. We find that the distribution takes its highest value at $c_0 = -23$, denoted by the blue point in this figure. The resulting N3LO pressure corresponding to this most-consistent value of $c_0$ is shown in the right panel of \cref{fig:p_N3LO_predict}, where the shaded region denotes the standard scale variation of $X \in [1/2,2]$. We find that for this value, the full N3LO pressure of cold QM lies inside the N2LO result and is well converged. In particular, the errors reach $\pm 25\%$ at $\muB = 2.2$~GeV, corresponding to about $n = 27 n_0$, denoted in the right panel of this figure by the dashed gray line. This is to be contracted with $n \approx 40 n_0$ for the N2LO result at $\muB = 2.6$~GeV. We note that between these densities, the relative importance of the pairing contribution to the pressure increases only by $\sim2$, so pairing effects do not grow appreciably. We thus see as an outcome of this Bayesian modeling that significant improvement to the cold-QM EOS may occur by computing the remainder of the N3LO pressure.

\emph{Conclusions.}---%
In this Letter, we have fully computed the contributions from the screened gluonic sector at next-to-next-to-next-to leading order in the strong coupling $\alpha_s$. This elevates the perturbative-QCD equation of state of cold quark matter to the $O(\alpha_s^3 \ln \alpha_s)$ level, leaving only one constant from the hard sector left to be computed, and finally places it on par with its high-temperature counterpart. We have achieved this by deriving the next-to-leading-order corrections to soft gluon propagation within a high-density medium, extending existing hard-thermal-loop results to the zero-temperature limit.

A natural follow-up to this work is the evaluation of the remaining hard constant. This would allow for the use of the full N3LO equation of state in applications, such as in studies of the neutron-star-matter equation of state. As this is an involved undertaking, it will require the development of new tools and methods. Some work in this direction has already been performed \cite{Ghisoiu:2016swa,Gorda:2022yex,Osterman:2023tnt} and more is ongoing. 

\begin{acknowledgments}
\emph{Acknowledgments.}---%
We thank Aleksi Kurkela, Aleksas Mazeliauskas, and Aleksi Vuorinen for their helpful comments and suggestions. This work is supported in part by the Deutsche Forschungsgemeinschaft (DFG, German Research Foundation)--Project ID 279384907--SFB 1245 and by the State of Hesse within the Research Cluster ELEMENTS (Project ID 500/10.006) (T.G.). R.P. and K.S. have been supported by the Academy of Finland Grants No.~347499 and No.~353772 as well as by the European Research Council, Grant No.~725369. S.S. acknowledges support of the DFG cluster of excellence ORIGINS funded by the DFG under Germany's Excellence Strategy - EXC-2094-390783311. In addition, K.S. gratefully acknowledges support from the Finnish Cultural Foundation.
\end{acknowledgments}

\bibliography{references}

\clearpage
\appendix

\begin{widetext}
\pagebreak
\section{Supplemental material}

Here we discuss some technical details of the computation of the gluon self-energy and the pressure computation presented in the Letter. We start by discussing the general structure of the soft gluon self-energy, to fix our notation, then present the new next-to-leading order (NLO) results for the soft gluon self-energies at large chemical potential and zero temperature $T$. We then continue to the calculation of the mixed diagrams, presented in the main text in \cref{eq:mixeddiags}. Finally, we discuss the general structure of the next-to-next-to-next-to-leading order (N3LO) pressure, including the cancellation of divergences between the different sectors, and we present results for general $\nc$ and $\nf$ quarks with equal chemical potentials $\muQ$\footnote{The generalization to unequal $\muQ$ is straightforward.} The results shown can be related to those of the main text simply by substituting $\muQ = \muB / 3$.

\subsection{Gluon self-energies}

At zero temperature and large density, the gluon self-energy tensor $\Pi_{\mu\nu}^{ab} \equiv \Pi_{\mu\nu}\delta^{ab}$ can be expanded in the QCD coupling $\as$ in the schematic form
\begin{equation}
    \Pi^{\mu\nu}(K)=\Pi_{1}^{\mu\nu}(K)+\Pi_{2}^{\mu\nu}(K)+O(\as^3)+\Pi_{\text{resum}}^{\mu\nu}(K),
\end{equation}
where the subscript counts the order in the naive loop expansion, and $\Pi_{\text{resum}}^{\mu\nu}$ encompasses all contributions involving (hard-thermal-loop) HTL-resummed diagrams, which we will ignore here.\footnote{They contribute to the N3LO pressure via HTL-resummed two-loop diagrams, already computed in \cite{Gorda:2021kme, Gorda:2021znl}, which contribute to $p^s$ in the main text.}  The one- and two-loop self-energies can further be expanded in small external Euclidean momentum $K=(K^0,\mathbf{k})$ as 
\begin{equation}\label{eq:SEsmallKexp}
    \Pi_{1}^{\mu\nu}(K)\equiv \mE^2 \Pi_{1,\text{HTL}}^{\mu\nu}(\hat{K}) + \frac{\as K^2}{4\pi}\Pi_{1,\text{Pow}}^{\mu\nu}(\hat{K}) + O(K^4), \quad  \Pi_{2}^{\mu\nu}(K)\equiv \frac{\as\mE^2}{4\pi}\Pi_{2,\text{HTL}}^{\mu\nu}(\hat{K}) + O(K^2),
\end{equation}
where $\hat{K} \equiv K/|K|$ and the self-energy coefficients $\Pi_{i,q}^{\mu\nu}$ have been made $\as$-independent and dimensionless by introducing the in-medium effective mass parameter $\mE$, which reads for $\nf$ identical quark flavors with equal chemical potentials $\muQ$ in $d \equiv D-1 = 3-2\epsilon$ dimensions \cite{Laine:2005ai,Gorda:2021kme}
\begin{equation}
\mE^2 \equiv \as\nf\muQ^2\left(\frac{e^{\gamE}\Lh^2}{4\pi\muQ^2}\right)^{\frac{3-d}{2}} \frac{4\Gamma(\frac{1}{2})}{(4\pi)^{\frac{d-1}{2}}\Gamma(\frac{d}{2})} = \frac{2N_f}{\pi} \as \muQ^2 + O(\varepsilon) 
\end{equation}
with $\Lh$ as the factorization scale and $\gamE$ as the Euler--Mascheroni constant.

The gluon self-energy is a symmetric rank-two tensor spanned by four basis tensors (for a comprehensive overview see \cite{Gorda:2023zwy}). However, for the small-$K$ coefficients shown in \cref{eq:SEsmallKexp}, only two of the four respective components turn out to be non-vanishing at $T=0$, namely the three-dimensionally transverse and longitudinal components, here denoted with subscripts T and L. The respective basis tensors are given by the projectors
\begin{equation}\label{eq:projTL}
    \mathbb{P}_\mathrm{T}^{\mu\nu}(\hat{K}) \equiv \delta_{i}^\mu\delta_{j}^\nu\left(\delta^{ij}-\frac{k^ik^j}{\mathbf{k}^2}\right), \quad \mathbb{P}_\mathrm{L}^{\mu\nu}(\hat{K})  \equiv \mathbb{P}^{\mu\nu}(\hat{K}) - \mathbb{P}_\mathrm{T}^{\mu\nu}(\hat{K}),
\end{equation}
where $\mathbb{P}^{\mu\nu}(\hat{K})=\delta^{\mu\nu}-\hat{K}^\mu\hat{K}^\nu$,  $i,j=1,\dots,d$, and $k^i$ being the components of $\mathbf{k}$. 
Further, the T and L components of the small-$K$ coefficients depend only on the Euclidean polar angle $\Phi$ through a variable $x$ defined as $x\equiv\tan\Phi=|\mathbf{k}|/K^0$. 

With the conventions laid out above, the previously known one-loop HTL self-energy components \cite{Braaten:1989mz,Manuel:1995td}, expanded to $O(\epsilon^2)$, read\footnote{For the $d$-dimensional expressions, see e.g. \cite{Gorda:2021kme}.}
\begin{align}
    \Pi^{1,\mathrm{HTL}}_\mathrm{T} &= \frac{\Ae}{2}+\frac{\Ae-\He}{2}\varepsilon+\frac{1}{2}\left\{1+\left(1-\frac{\pi^2}{6}-2\ln 2+2\ln^2 2\right)\Be+\left(1-2\ln 2\right)\He-\Je\right\}\varepsilon^2+O(\varepsilon^3), \\
    \Pi^{1,\mathrm{HTL}}_\mathrm{L} &= \Be+\He\varepsilon-\left\{\left(2-\frac{\pi^2}{6}-2\ln 2+2\ln^2 2\right)\Be+2\left(1-\ln 2\right)\He-\Je\right\}\varepsilon^2+O(\varepsilon^3),
\end{align}
where we have used the shorthands
\begin{equation}
\Le \equiv \frac{1}{x}\arctan(x), \qquad \Ae \equiv \left(1+\frac{1}{x^2}\right)\Le - \frac{1}{x^2}, \qquad \Be \equiv \left(1+\frac{1}{x^2}\right)\left (1-\Le \right ),    
\end{equation}
and
\begin{align}
\He&\equiv\left(1+\frac{1}{x^2}\right)\left\{ 2+\ln\left[\frac{1}{4}\left(1+\frac{1}{x^{2}}\right)\right]\right\} \Le-\frac{i}{2x}\left[\mathrm{Li}_{2}\left(\frac{i+x}{i-x}\right)-\mathrm{Li}_{2}\left(\frac{i-x}{i+x}\right)\right], \\
\begin{split}
\Je &\equiv \left(1+\frac{1}{x^2}\right)\biggl\{2-2\ln 2+2\ln^2 2-\frac{\pi^2}{6}+\biggl[1-\ln 2-\ln^2 2-\frac{\pi^2}{12}+ix\ln\left(\frac{x}{i+x}\right)\Le+\frac{x^2\Le^2}{3}\biggr]2\Le \\
&\quad+\frac{i}{x}\biggl[(2\ln 2)\mathrm{Li}_2\left(\frac{2x}{i+x}\right)-2\mathrm{Li}_3\left(\frac{2x}{i+x}\right)-\mathrm{Li}_3\left(\frac{i-x}{i+x}\right)+\zeta(3)\biggr]\biggr\}.
\end{split}
\end{align}
Here $\mathrm{Li}_s$ and $\zeta$ are the standard polylogarithm and (Riemann) zeta functions respectively. We also note that the standard one-loop resummed HTL-propagator \cite{Braaten:1989mz,Manuel:1995td} (still with the trivial color indices suppressed) reads
\begin{equation}
    G_\lo^{\mu\nu}(K)=\frac{\mathbb{P}_\text{T}^{\mu\nu}(\hat{K})}{K^2+\mE^2\Pi^{1,\text{HTL}}_{\text{T}}(\Phi)} + \frac{\mathbb{P}_\text{L}^{\mu\nu}(\hat{K})}{K^2+\mE^2\Pi^{1,\text{HTL}}_{\text{L}}(\Phi)} + \xi \frac{\hat{K}^\mu \hat{K}^\nu}{K^2} 
\end{equation}
with $\xi$ being the gauge parameter.

The self-energies $\Pi_{2,\text{HTL}}^{\mu\nu}$ and $\Pi_{1,\text{Pow}}^{\mu\nu}$ have recently been evaluated at nonzero $T$ and $\muQ$ in general covariant $\xi$-gauge \cite{Gorda:2023zwy} (and in Feynman gauge $\xi =1$ in \cite{Ekstedt:2023oqb}). Here, we present new results for these gluon self-energies at exactly $T=0$, as those in \cite{Gorda:2023zwy} contained problematic terms proportional to $\ln T$, which have now been converted into infrared $1/\epsilon$ divergences. In order to reach the $T=0$ limit, one has to consider the intermediate integral expressions leading to the self-energy: The problematic terms arise from bosonic integrals of the form $\int_0^\infty \mathrm{d}p \,p^{d-4} n_B(p)$, with $n_B$ the standard Bose--Einstein distribution function. They lead to infrared divergences proportional to $(T^2)^{\varepsilon}$ if the limit $T\rightarrow 0$ is taken after integration, but vanish due to being scale-free when computed at $T=0$. In order to get the correct $T=0$ self-energies, we isolate these integrals and set them to zero, leading to a final expression that is free from $\ln T$-type singularities. Therefore, the $d$-dimensional bare result for the two-loop self-energy reads
\begin{align}
\label{eq:2HTLTd}
    \begin{split}
    \Pi^{2,\mathrm{HTL}}_\mathrm{T,bare} &= \frac{\normf\cf}{(d-1)(d-3)}\left(1+\frac{1}{x^2}\right)\biggl\{\frac{1-(d-2)x^2}{1+x^2}(d-2)\mathcal{A}_0+2i(d-3)\mathcal{A}_1 \\
    &\quad-\left[(d-6)\left(1+x^2\right)+4x^2\right]\mathcal{A}_2+2i\left(1+x^2\right)\mathcal{A}_3\biggr\},
    \end{split} \\
\label{eq:2HTLLd}
    \begin{split}
    \Pi^{2,\mathrm{HTL}}_\mathrm{L,bare} &= \normf\left(1+\frac{1}{x^2}\right)\biggl\{(\ca-2\cf)\biggl[\frac{d-1}{2(d-2)^2\mathcal{A}_0}\left(\left(1+x^2\right)\mathcal{A}_2-i\mathcal{A}_1\right)^2\biggr] \\
    &\quad-\frac{\cf}{d-3}\biggl[(d-2)\mathcal{A}_0+2i(d-3)\mathcal{A}_1-\left((d-6)\left(1+x^2\right)+2x^2\right)\mathcal{A}_2+2i\left(1+x^2\right)\mathcal{A}_3\biggr]\biggr\},
    \end{split}
\end{align}
where we have defined the prefactor $\normf\equiv(e^{\gamE/2}\Lbar/\muQ)^{2\varepsilon}/\Gamma(\frac{d-1}{2})$ and the integral
\begin{equation}
    \mathcal{A}_\alpha\equiv\int_{-1}^1\mathrm{d}z\frac{(1-z^2)^\frac{d-3}{2}}{\left(-i+xz\right)^\alpha}
\end{equation}
which can be expressed in a closed form in terms of the Gauss hypergeometric function.  The bare result becomes finite upon renormalization and can be expanded in $\varepsilon$ yielding:
\begin{align}
\label{eq:2HTLT}
    \Pi^{2,\mathrm{HTL}}_\mathrm{T} &= \frac{\ca(\xih-4)}{2\varepsilon}\Pi^{1,\mathrm{HTL}}_\mathrm{T}-2\cf\Le-2\cf\left\{2\left(1+\ln\frac{\Lbar}{2\muQ}\right)\Le-\frac{x^2}{1+x^2}\He\right\}\varepsilon+O(\varepsilon^2), \\
\label{eq:2HTLL}
    \begin{split}
    \Pi^{2,\mathrm{HTL}}_\mathrm{L} &= \frac{\ca(\xih-4)}{2\varepsilon}\Pi^{1,\mathrm{HTL}}_\mathrm{L}+2(\ca-2\cf)(1-\Le)\Be-2\cf \\
    &\quad+\left\{2(\ca-2\cf)\left[(1-\Le)\left(1+2\ln\frac{\Lbar}{2\muQ}\right)+\frac{2x^2}{1+x^2}\He\right]\Be-4\cf\left(\ln\frac{\Lbar}{2\muQ}+\Le\right)\right\}\varepsilon+O(\varepsilon^2).
    \end{split}
\end{align}
For the $d$-dimensional bare power correction, we have:
\begin{align}
\label{eq:1PowTd}
    \Pi^{1,\mathrm{Pow}}_\mathrm{T,bare} &= \frac{\normf\nf}{(d-1)(d-3)}\left(1+\frac{1}{x^2}\right)\biggl\{\left[(d-2)x^2-1\right]\mathcal{A}_2-2i\left(1+x^2\right)\mathcal{A}_3+\left(1+x^2\right)^2\mathcal{A}_4\biggr\}, \\
\label{eq:1PowLd}
    \Pi^{1,\mathrm{Pow}}_\mathrm{L,bare} &= \frac{\normf\nf}{d-3}\left(1+\frac{1}{x^2}\right)\left(1+x^2\right)\biggl\{\mathcal{A}_2+2i\mathcal{A}_3-\left(1+x^2\right)\mathcal{A}_4\biggr\}.
\end{align}
The corresponding renormalized and $\varepsilon$-expanded expressions read
\begin{align}
\label{eq:1PowT}
    \begin{split}
    \Pi^{1,\mathrm{Pow}}_\mathrm{T} &= \frac{\ca(10+3\xih)}{6\varepsilon}+\frac{\nf}{3}\left(4\ln\frac{\Lbar}{2\muQ}+4\Le-\Ae\right) \\
    &\quad-\frac{\nf}{3}\left\{\frac{\pi^2}{2}+2(\Ae-4\Le)\ln\frac{\Lbar}{2\muQ}-4\ln^2\frac{\Lbar}{2\muQ}-6\Le-\frac{1-3x^2}{1+x^2}\He\right\}\varepsilon+O(\varepsilon^2),
    \end{split} \\
\label{eq:1PowL}
    \begin{split}
    \Pi^{1,\mathrm{Pow}}_\mathrm{L} &= \frac{\ca(10+3\xih)}{6\varepsilon}+\frac{2\nf}{3}\left(2\ln\frac{\Lbar}{2\muQ}+2\Le-\Be\right) \\
    &\quad-\frac{\nf}{3}\left\{2+\frac{\pi^2}{2}+4(\Be-2\Le)\ln\frac{\Lbar}{2\muQ}-4\ln^2\frac{\Lbar}{2\muQ}-2\Be-8\Le+\frac{2+6x^2}{1+x^2}\He\right\}\varepsilon+O(\varepsilon^2),
    \end{split}
\end{align}
where $\xih \equiv 1-\xi$, so that $\xih=0$ corresponds to Feynman gauge and $\xih=1$ corresponds to Landau gauge. We note that there are no terms proportional to $C_A$ present in the $O(\epsilon^0)$ and $O(\epsilon)$ contributions to $\Pi^{1,\mathrm{Pow}}$. This occurs because at $T=0$ all gluonic diagrams contributing to the power correction are scaleless and therefore vanish in dimensional regularization. The exception to this is the renormalization counterterm, which shows up as the vacuum divergence $\propto C_A/\varepsilon$ present above.

\subsection{Mixed diagrams}

In this subsection, we detail the evaluation of the mixed diagrams in \cref{eq:mixeddiags} 
using the self-energies in 
\cref{eq:2HTLT,eq:2HTLL,eq:1PowT,eq:1PowL}. 
We start by writing \cref{eq:mixeddiags} in Euclidean spherical variables as\footnote{Note that the resummed propagator and the self-energies alone also depend on $\hat{K}$ through the projectors [see \cref{eq:projTL}], but the dependence vanishes upon contracting the tensors and taking the trace.}
\begin{equation}\label{eq:p3m1}
    p^m = -\frac{\as\mE^2\da}{8\pi}\int_K\tr{G_\lo(|K|,\Phi)\left[\Pi^{2,\mathrm{HTL}}(\Phi)+\frac{K^2}{\mE^2}\Pi^{1,\mathrm{Pow}}(\Phi)\right]} 
\end{equation}
with the integration measure
\begin{equation}
    \int_K \equiv \left (\frac{e^{\gamE}\Lh^2}{4\pi} \right )^{\frac{3-d}{2}} \frac{(4\pi)^{-d/2}}{\Gamma(d/2)\pi} \int_0^\pi \ud \Phi \sin^{d-1}(\Phi)\int_0^\infty \ud |K| |K|^d.
\end{equation}
The radial integral in \cref{eq:p3m1} can be performed by using the result
\begin{equation}
\label{eq:radialint}
\int_0^\infty\frac{\ud |K||K|^a}{K^2+\mE^2\Pi^{1,\mathrm{HTL}}_{\rm I}(\Phi)} = \frac{\pi}{2}\left(\mE^2\Pi^{1,\mathrm{HTL}}_{\rm I}(\Phi)\right)^{\frac{a-1}{2}}\sec\left(\frac{a\pi}{2}\right),
\end{equation}
for $\rm{I} \in \{\rm{T}, \rm{L} \} $, yielding
\begin{equation}
\label{eq:p3m2}
\begin{split}
p^m = -\frac{\as\mE^4\da}{8(4\pi)^{2}} \left (\frac{e^{\gamE}\Lh^2}{\mE^2} \right )^{\frac{3-d}{2}} \frac{\sec\left ( \frac{\pi d}{2}\right )}{\Gamma(\frac{d+1}{2})}  \biggl \langle \tr{\left (\Pi^{1,\text{HTL}}\right )^{\frac{d-1}{2}}\biggl [\Pi^{2,\text{HTL}}-\Pi^{1,\text{HTL}}\Pi^{1,\text{Pow}}\biggr ]} \bigg\rangle_d,
\end{split}
\end{equation}
where we use the angle brackets $\langle \cdot \rangle_d$ to denote a normalised angular average in $d$ spatial dimensions. In particular, we are interested in functions $f$ which depend only on the angle $\Phi$, in which case 
\begin{equation}
\biggl \langle f(\Phi) \bigg\rangle_d \equiv  \frac{\int_0^\pi \ud \Phi \sin^{d-1}(\Phi) f(\Phi)}{\int_0^\pi \ud \Phi \sin^{d-1}(\Phi)}.
\end{equation}
Note that the radial $|K|$ integral resulted in the combination $\Pi^{2,\text{HTL}} - \Pi^{1,\text{HTL}}\Pi^{1,\text{Pow}}$ in which the (modified) gauge parameter $\hat\xi$ cancels out, demonstrating the explicit gauge invariance of the mixed sector.

The remaining angular average expanded in $\varepsilon$ yields
\begin{equation}\label{eq:angavg}
\biggl \langle \tr{\left (\Pi^{1,\mathrm{HTL}}\right )^{1-\varepsilon}\biggl [\Pi^{2,\mathrm{HTL}}-\Pi^{1,\mathrm{HTL}}\Pi^{1,\mathrm{Pow}}\biggr ]} \bigg\rangle_{3-2\varepsilon} = \frac{\mathcal{A}_{-1}}{2\varepsilon}  + \mathcal{A}_0 + \mathcal{A}_1 (2\varepsilon) + O(\varepsilon^2),   
\end{equation}
where the leading $1/(2\varepsilon)$-coefficient is given by the following analytical result 
\begin{equation}
\begin{split}
\mathcal{A}_{-1} = -\frac{22}{3}\biggl\langle\frac{\Ae^2}{2}+\Be^2\biggr\rangle_3 \ca = -\frac{11}{3}\ca.
\end{split}
\end{equation}
The remaining two subleading coefficients contain numerical parts as well, reading
\begin{equation}
    \mathcal{A}_{0}=-1.01478\ca-\left(11-\frac{5\pi^{2}}{6}\right)\cf+\left(\frac{5}{12}-\frac{7\pi^{2}}{144}\right)\nf-\frac{2}{3}\nf\ln\frac{\Lbar}{2\muQ},
\end{equation}
and
\begin{equation}
\begin{split}
    \mathcal{A}_{1}&=-0.0718467\ca+\left(3-\frac{\pi^2}{4}\right)\ca\ln\frac{\Lbar}{2\muQ}-2.41110\cf-\left(11-\frac{5\pi^{2}}{6}\right)\cf\ln\frac{\Lbar}{2\muQ}\\
    &\quad+0.100489\nf-0.344447\nf\ln\frac{\Lbar}{2\muQ}-\frac{1}{3}\nf\ln^2\frac{\Lbar}{2\muQ}.
\end{split}
\end{equation}

\subsection{Structure of the pressure}
As explained in the main text and in extensive detail in \cite{Gorda:2021kme,Gorda:2021znl}, the N3LO pressure can be organized into three distinct sectors---soft, hard, and mixed---according to the kinematics of the two independent gluon lines as in \cref{eq:pressectors}. The soft sector arises from a kinematic region where both gluons are soft, diagrammatically corresponding to two-loop diagrams in the HTL theory. In dimensional regularization, with $d=3-2\varepsilon$ spatial dimensions, the result for the soft sector has the form
\begin{equation}
\begin{split}\label{eq:p3soft}
p^s & = 
\frac{\as\mE^4\da}{(4\pi)^3} \left(\frac{\mE}{\Lh}\right)^{-4\varepsilon}\left(\frac{p^s_{-2}}{(2\varepsilon)^2} + \frac{p^s_{-1}}{2\varepsilon} + p^s_{0} \right),
\end{split}
\end{equation}
where the coefficients $p_i^s$ can be read off from \cite{Gorda:2021znl} and for the sake of completeness they are also given below.\footnote{Note the factor of $\pi\nc$ difference in the definition of $p_i^s$ from that work.} On the other hand, the hard sector stems from the region where both gluons are hard, corresponding to four-loop diagrams within the naive loop expansion. The renormalized result reads schematically as
\begin{equation}
\begin{split}\label{eq:p3hard}
p^h & = 
\frac{\as\mE^4\da}{(4\pi)^3}\left(\frac{\muQ}{\Lh}\right)^{-4\varepsilon}\left(\frac{p^h_{-2}}{(2\varepsilon)^2} + \frac{p^h_{-1}(\Lbar)}{2\varepsilon} + p^h_{0}(\Lbar) \right),
\end{split}
\end{equation}
with the coefficients $p_i^h$ currently unknown. Finally, the mixed sector corresponds to the region where one of the gluons is soft while the other remains hard giving partially HTL-resummed three-loop diagrams (see the main text). The renormalized mixed result has the form
\begin{equation}
\begin{split}\label{eq:p3mixed}
p^m  = 
\frac{\as\mE^4\da}{(4\pi)^3} \left(\frac{\mE}{\Lh}\right)^{-2\varepsilon}\left(\frac{\muQ}{\Lh}\right)^{-2\varepsilon}\left(\frac{p^m_{-2}}{(2\varepsilon)^2} + \frac{p^m_{-1}(\Lbar)}{2\varepsilon} + p^m_{0}(\Lbar) \right).
\end{split}
\end{equation}
As a result of this Letter, the coefficients $p_i^m$ are given in the main text in \cref{eq:p3results} for $\nc=\nf=3$ and below for general $\nc$ and $\nf$.

With the angular average in \cref{eq:angavg} known to $O(\varepsilon)$, it can be inserted into \cref{eq:p3m2}, which, expanded in $\varepsilon$ and truncating at the constant order, yields the mixed coefficients $p_{i}^m$. Note that in order to write \cref{eq:p3m2} in the form of \cref{eq:p3mixed} we introduce unity as the factor $1=\left(\muQ/\Lh\right)^{-2\varepsilon}\left(\muQ/\Lh\right)^{2\varepsilon}$, where the first half is absorbed into the prefactors shown in \cref{eq:p3mixed} and the second half is expanded and the scales $\Lh$ and $\Lbar$ are set equal. The result for the coefficients $p_i^m$ then reads
\begin{equation}
    \begin{split}\label{eq:mixedcoef}
        p^m_{-2} & = -\frac{11\ca}{3}, \\
        p^m_{-1} & = \frac{11}{3}\left(\ca-\frac{2\nf}{11}\right)\ln \frac{\Lbar}{2\muQ}-0.306569\ca-\left(11-\frac{5\pi^2}{6}\right)\cf+\left(\frac{5}{12}-\frac{7\pi^2}{144}\right)\nf, \\
        p^m_0 & = -\frac{11}{6}\left(\ca-\frac{2\nf}{11}\right)\ln^2 \frac{\Lbar}{2\muQ} +\left(0.839168\ca-0.152576\nf\right)\ln \frac{\Lbar}{2\muQ}\\
        &\quad-1.15650\ca-1.87506\cf+0.112677\nf.
    \end{split}
\end{equation}

On the other hand, the soft coefficients to the pressure from \cite{Gorda:2021znl} can be converted to the conventions used here and read
\begin{equation}\label{eq:softcoef}
p^s_{-2} = \frac{11\ca}{6}, \qquad p^s_{-1} = 4.73535(60)\cdot\ca, \qquad p^s_0 = 6.9508(28)\cdot\ca.
\end{equation}

Throughout this supplemental material, numerical results arise in averages involving higher-order $\varepsilon$-expansions of the self-energies and measures, such as logarithms of the self-energy. While partial analytic contributions to these numerical results can be extracted, we have opted against doing so as the resulting forms are not particularly illuminating. The fact that these integrals do not appear to possess a closed-form representation does not limit precision in a meaningful way (unlike in the contributions to the soft sector \cite{Gorda:2021kme}), as the integrals are one-dimensional and can be performed to machine precision with standard numerical integration packages.

With the soft and mixed divergences known, the hard divergences are also immediately determined by them, as the sum of the divergences must cancel if HTL is the correct theory of soft gluons, i.e., the coefficients should satisfy
\begin{equation}\label{eq:divrelation}
p^s_{-2}+p^m_{-2}+p^h_{-2}=0, \qquad p^s_{-2}=p^h_{-2}, \qquad p^s_{-1}+p^m_{-1}+p^h_{-1}=0,
\end{equation}
in order to cancel the divergences in the sum $p_{\text{N3LO}}=p^s+p^m+p^h$. Thus, we immediately infer
\begin{equation}
    \begin{split}\label{eq:hardcoef1}
        p^h_{-2} & = \frac{11\ca}{6}, \\
        p^h_{-1} & = -\frac{11}{3}\left(\ca-\frac{2\nf}{11}\right)\ln \frac{\Lbar}{2\muQ}-4.42878(60)\cdot\ca-\left(\frac{5\pi^2}{6}-11\right)\cf-\left(\frac{5}{12}-\frac{7\pi^2}{144}\right)\nf. 
    \end{split}
\end{equation}
The remaining finite hard term $p_0^h$ can be partially determined from renormalization-scale independence of the complete pressure (see \cite{Gorda:2022zyc}, the argument holds also in non-Abelian theory):
\begin{equation}
\begin{split}\label{eq:hardcoef2}
        p^h_0 & = \frac{1}{36}\left(110\ca-\frac{121\ca^2}{\nf}-16\nf\right)\ln^2 \frac{\Lbar}{2\muQ} +\biggl(4.39077\ca-\frac{7}{6}\cf-\frac{1715\ca^2}{108\nf}\\
        &\quad+\frac{187\ca\cf}{24\nf}-0.244450\nf\biggr)\ln \frac{\Lbar}{2\muQ}+c_0,
\end{split}
\end{equation}
where the $\Lbar$-independent constant $c_0$ arises purely from the IR-finite parts of (potentially IR-divergent) four-loop unresummed diagrams. Finally, it is worth noting that \cref{eq:divrelation} implies $p_{-2}^m=-2p_{-2}^s$, which our calculation now explicitly confirms (see \cref{eq:mixedcoef,eq:softcoef}).

The full N3LO contribution to the pressure, as in \cref{eq:p3unknowns} in the main text, is obtained by summing the soft, mixed, and hard sectors in \cref{eq:p3soft,eq:p3hard,eq:p3mixed} together and plugging in the coefficients determined above. This leads to the cancellation of both the $1/\varepsilon$ divergences and the dependence on the factorization scale $\Lh$ as expected. Including the lower-order terms, the result for the pressure of cold and dense QM can be written in the form
\begin{equation}
\begin{split}
\frac{p}{p_\text{free}}&= 1 + \left(\frac{\as}{\pi}\right)a_{1,1}+\nf\left(\frac{\as}{\pi}\right)^2\left[a_{2,1}\ln\left(\nf\frac{\as}{\pi}\right)+a_{2,2}\ln \frac{\Lbar}{2\muQ}+a_{2,3}\right] \\
     &\quad+\nf^2\left(\frac{\as}{\pi}\right)^3\biggl[a_{3,1}\ln^2\left(\nf\frac{\as}{\pi}\right)+a_{3,2}\ln\left(\nf\frac{\as}{\pi}\right)+a_{3,3}\ln\left(\nf\frac{\as}{\pi}\right)\ln\frac{\Lbar}{2\muQ} \\
     &\quad+a_{3,4}\ln^2\frac{\Lbar}{2\muQ}+a_{3,5}\ln\frac{\Lbar}{2\muQ}+a_{3,6} \biggr]+O(\alpha_s^4), \label{eq:presform}
\end{split}
\end{equation}
where the coefficients $a_{i,j}$ are given in \cref{tab:pressurecoeffs}, $p_\text{free}=\muQ^4 \nf \nc /(12\pi^2)$, and $\muQ$ is the quark chemical potential, taken here to be equal for each of the $\nf$ flavors.

\begin{table}[t!]
\begin{tabular}{@{\quad}l@{\quad\quad}l@{\quad\quad}l}
\toprule
    $a_{1,1}$ &   $-\frac{3}{4}\da\nc^{-1}$ & $-2$ \\
    $a_{2,1}$ &   $-\frac{3}{8}\da\nc^{-1}$ & $-1$ \\
    $a_{2,2}$ &   $b_0 a_{1,1}$ & $-3$ \\
    $a_{2,3}$ &   $\da\Bigl\{\left(\frac{11}{6}-\frac{\pi^2}{8}-\frac{5}{4}\ln 2+\ln^2 2-\frac{3}{16}\delta\right)\nc^{-1}$ & $-5.0021$ \\
    &\quad$-\frac{415}{192}\nf^{-1}-\frac{51}{64}\nc^{-2}\nf^{-1}\Bigr\}$ & \\
    $a_{3,1}$ &   $\frac{11}{32}\da\nf^{-1}$ & $\frac{11}{12}$ \\
    $a_{3,2}$  &  $\da\Bigl(0.023665\nc^{-1}-2.4396(5)\cdot\nf^{-1}$ & $-6.5968(12)$ \\
    &\quad$-0.52037\nc^{-2}\nf^{-1}\Bigr)$ & \\
    $a_{3,3}$ &   $2b_0a_{2,1}$ & $-3$ \\
    $a_{3,4}$  &  $b_0^2a_{1,1}$ & $-\frac{9}{2}$ \\
    $a_{3,5}$  &  $b_0a_{2,1}+2b_0a_{2,3}+b_1a_{1,1}$ & $-18.284$ \\
    $a_{3,6}$  &  $\da\Bigl(0.10091\nc^{-1}+1.7864(18)\cdot\nf^{-1}$ & $5.1342(48) + \frac{2}{3} c_0$ \\
    &\quad$+0.34245\nc^{-2}\nf^{-1}+\frac{3}{4}c_0\nc^{-1}\nf^{-1}\Bigr)$ & \\
\botrule
\end{tabular}
    \caption{List of numerical values for the coefficients appearing in \cref{eq:presform}. The rightmost column shows the numerical values of the coefficients for $\da=8$, $\nc=3$, and $\nf=3$. The beta-function coefficients $b_i$ are given by $b_0=\frac{11}{6}\nc\nf^{-1}-\frac{1}{3}$ and $b_1=\frac{17}{12}\nc^2\nf^{-2}+\frac{1}{8}\nc^{-1}\nf^{-1}-\frac{13}{24}\nc\nf^{-1}$. The $N_c,N_f$-independent constant $\delta$ is defined in \cite{Vuorinen:2003fs} and has the numerical value $\delta=-0.85638$ up to 5-digit precision.} 
    \label{tab:pressurecoeffs}
\end{table}

\end{widetext}

\end{document}